\documentclass[aps,prl,twocolumn,footinbib,longbibliography]{revtex4-1}
\usepackage{amsmath}
\usepackage{amsfonts}
\usepackage{amssymb}
\usepackage{lmodern,dsfont}
\usepackage{graphicx}
\usepackage[usenames,dvipsnames]{xcolor}
\usepackage{bm}
\usepackage{blkarray}
\usepackage{blindtext}
\usepackage[english=american]{csquotes}

\newcommand{\kb}{k_\text{B}}
\newcommand{\Av}[1]{\left\langle #1 \right\rangle}
\newcommand{\av}[1]{\langle #1 \rangle}
\newcommand{\n}{\nonumber}

\begin{document}

\author{Andreas Dechant}
\affiliation{WPI-Advanced Institute of Materials Research (WPI-AIMR), Tohoku University, Sendai 980-8577, Japan}
\title{Estimating the diffusion coefficient of trapped particles}
\date{\today}

\begin{abstract}
We show that observing the trajectories of confined particles in a thermal equilibrium state yields an estimate on the free-space diffusion coefficient.
For generic trapping potentials and interactions between particles, the estimate comes in the form of a lower bound on the true diffusion coefficient.
For non-interacting particles in harmonic trapping potentials, which approximately describes many experimental situations, the estimate is asymptotically exact.
This allows to determine the diffusion coefficient from an equilibrium measurement, as opposed to a direct observation of diffusion, which necessarily starts from a non-equilibrium state.
We explicitly demonstrate that the estimate remains quantitatively accurate in the presence of weak interactions and anharmonic corrections.
\end{abstract}

\maketitle

Diffusion of particles in contact with a heat bath is a direct, macroscopic consequence of the microscopic thermal motion \cite{Ein05,Sut05,Smo06}. 
Consequently, a measurement of the diffusion coefficient allows access to microscopic properties of the environment such as the particle mobility via the Einstein relation \cite{Ein05}. 
However, in any experiment, the size of the system is finite and diffusion can only be observed transiently; it necessarily starts from a non-equilibrium initial condition. 
This can make a direct measurement of the diffusion coefficient difficult or even impossible.
For the paradigmatic example of diffusion, a cloud of small particles immersed in water inside a circular Petri dish, the particles should initially be inside a small area near the center of the dish, as to eliminate any influence of the walls.
On the other hand, they should not be too tightly concentrated, because otherwise the interactions between the particles cannot be neglected.
Finally, the Petri dish needs to be sufficiently large to allow for ample time to observe the spreading of the cloud before the particles are evenly distributed across the dish.
While these conditions are easy enough to satisfy in the classical tabletop experiment, it is not hard to imagine experimental situations where this is not possible. 
If the system size is small, the available time to observe diffusion may be too short for an accurate estimate of the diffusion coefficient, or it may be difficult to prepare a well-defined initial non-equilibrium state with the observed particle sufficiently far from the walls to exclude interactions.
Even in approximately unconfined situations, the ability to observe diffusion might be limited simply by particles diffusing out of the field of view of the microscope.

By contrast, equilibrium measurements of trapped particles face none of these restrictions.
Since, by definition, the state of the system does not change over time in equilibrium, the measurement time can be arbitrarily long.
No special care needs to be taken in the preparation of the initial state, since the equilibrium state is realized automatically by just waiting.
One may also adjust the confinement to keep the particles inside the field of view of the optical detection system.
However, the macroscopic equilibrium state by itself yields no information about the diffusive dynamics.
Nevertheless, even in the equilibrium state, the thermal motion at the micro-scale does not stop.
The emergence of single particle tracking experiments has made it feasible to investigate the motion of individual, microscopic particles in the macroscopic equilibrium state \cite{Man15}. 
It is suggestive that such measurements should be able to yield information on the microscopic properties of the system.
However, it is generally not straightforward to connect the trajectory data obtained from measurements to specific microscopic parameters.

Here we discuss a possibility to establish such a connection for confined particles in contact with a heat bath.
We show that, independent of the precise details of the system, it is always possible to obtain a lower bound on the free-space diffusion coefficient of individual particles using only a measurement of the particle trajectory, without any knowledge about the microscopic details of the system.
For non-interacting particles in a harmonic confinement, the lower bound becomes an equality.
In this situation, which to a good approximation describes many experimental situations, our result thus enables a direct determination of the free-space diffusion coefficient from a measurement of the trajectories of confined particles.
For anharmonic confining forces or interactions between particles, our estimate provides a lower bound on the diffusion coefficient.
As we demonstrate by numerical simulations, the estimate is quantitatively accurate even in the presence of moderately strong anharmonic corrections and interactions, provided that the dynamics of the particles is dominated by the interaction with the environment.

\textit{Model and bound.}
We consider $M$ Brownian particles immersed in an environment that is characterized by the temperature $T$ and the mobility $\mu$.
The motion of the particles is described by the coupled Langevin equations ($i = 1,\ldots,M$) \cite{Ris86,Cof04}
\begin{align}
\dot{x}_i(t) = -\mu \partial_{x_i} U(\bm{x}(t)) + \sqrt{2 \mu T} \xi_i(t) \label{langevin},
\end{align}
where we set $\kb = 1$ for convenience of notation.
The potential $U(\bm{x})$ includes the external confining potential, but also interactions between the particles, and the vector $\bm{x}(t) = (x_1(t),\ldots,x_M(t))$ encodes the positions of the particles.
The noises $\xi_i(t)$ are mutually independent, Gaussian and white, $\av{\xi_i(t) \xi_j(s)} = \delta_{i j} \delta(t-s)$.
We assume that the potential admits a well-defined equilibrium state
\begin{align}
P^\text{eq}(\bm{x}) = \frac{e^{-\frac{U(\bm{x})}{T}}}{Z} \quad \text{with} \quad Z = \int d\bm{x} \ e^{-\frac{U(\bm{x})}{T}} \label{boltzmann-gibbs} .
\end{align}
In the absence of the potential, $U = 0$, each particle undergoes Brownian motion, with the mean-square displacement growing as \cite{Ein05,Sut05,Smo06}
\begin{align}
\Av{\Delta x_i^2}_t = 2 D t \qquad \text{with} \qquad D = \mu T \label{diffusion} .
\end{align}
The Einstein relation between the diffusion coefficient $D$ and the mobility $\mu$ allows to obtain $\mu$ from a measurement of the diffusion coefficient.
In the presence of the potential, the equilibrium state Eq.~\eqref{boltzmann-gibbs} is independent of $D$ and $\mu$ and thus provides no information about the latter.
Our goal is to determine the free-space diffusion coefficient $D$ from a measurement of the trajectory $x_i(t)$ in the presence of the potential.
The method we propose relies on a measurement of the time-averaged position of the particle
\begin{align}
\bar{x}_i(t) = \frac{1}{t} \int_0^t ds \ x_i(s) .
\end{align}
If the system is ergodic, then this time average will coincide with the ensemble average for long measurement times, $\lim_{t \rightarrow \infty} \bar{x}_i(t) = \av{x_i}^\text{eq}$, where $\av{\ldots}^\text{eq}$ is an average with respect to the equilibrium state Eq.~\eqref{boltzmann-gibbs} \cite{Pap02}.
However, for any finite time, the time average is a random quantity, characterized by a finite variance $\langle \Delta \bar{x}_i^2 \rangle_t = \langle \bar{x}_i^2 \rangle_t - \langle \bar{x}_i \rangle^2_t$.
Our main result is the following inequality for the free-space diffusion coefficient,
\begin{align}
D \geq D^* \equiv \lim_{t \rightarrow \infty} \frac{2 \big(\av{\Delta x_i^2}^\text{eq}\big)^2}{t \av{\Delta \bar{x}_i^2}_t} \label{diffusion-bound}.
\end{align}
We can thus bound the diffusion coefficient from below by a lower estimate $D^*$, obtained by measuring the variances of the position and the time-averaged position in the equilibrium state.
Importantly, the right hand side is independent of any details of the potential or the environment and can be evaluated by measuring only the trajectory of the trapped particle in the equilibrium state.
Further, as discussed below, the lower estimate $D^*$ of the diffusion coefficient given by Eq.~\eqref{diffusion-bound} is valid irrespective of the overdamped limit and can also be applied to underdamped systems.

\textit{Non-interacting particles.}
If there is no interaction between the particles, Eq.~\eqref{langevin} reduces to a one-particle Langevin equation.
It was shown in Ref.~\cite{Dec11} that the fluctuations of the time average decay as
\begin{align}
&\av{\Delta \bar{x}^2}_t \simeq \frac{2}{\mu T Z t} \label{ta-variance} \\
&\qquad \times \int_{-\infty}^{\infty} dx \ e^{\frac{U(x)}{T}} \bigg[ \int_x^\infty dy \ (y-\av{x}^\text{eq}) \ e^{-\frac{U(y)}{T}} \bigg]^2 \n
\end{align}
for long times.
We remark that using this explicit expression, the bound \eqref{diffusion-bound} can be derived by applying the Cauchy-Schwarz inequality to the above integral.
For the particular case of an harmonic potential $U(x) = k x^2/2$ with spring constant $k$, the integral can be evaluated explicitly,
\begin{align}
\Av{\Delta \bar{x}^2}_t \simeq \frac{2 T}{\mu k^2 t} = \frac{2 \big( \av{\Delta x^2}^\text{eq} \big)^2}{\mu T t} .
\end{align}
Plugging this into the bound Eq.~\eqref{diffusion-bound}, we obtain
\begin{align}
D^* = \mu T \label{diffusion-bound-harmonic} .
\end{align}
Since the right-hand side is just equal to $D$ via Eq.~\eqref{diffusion}, we see that equality in Eq.~\eqref{diffusion-bound} holds if the confining potential is harmonic.
For anharmonic confining potentials, the bound \eqref{diffusion-bound} provides a lower estimate on the diffusion coefficient. 
For deep potentials $U \gg T$, a harmonic approximation works well in most cases and thus the error of the estimate is expected to be small in such cases.
However, this is obviously only true as long as the particle remains close to a minimum of the potential; for example, barrier crossing events are not captured by the harmonic approximation and generally spoil the estimate since they lead to large fluctuations of the time average.
The identity \eqref{diffusion-bound-harmonic} suggests a new way to measure the diffusion coefficient of non-interacting particles:
Instead of observing their free-space diffusion directly, with its associated limitations of finite measurement time and preparation of the initial state, one may also trap the particles inside a harmonic confinement and observe their trajectories in the trapped equilibrium state.
Then the right-hand side of Eq.~\eqref{diffusion-bound} can be used to determine the diffusion coefficient exactly in the long-time limit.

\textit{Interacting particles.}
In the case of interacting particles, there is generally no explicit expression for the variance of the time average similar to Eq.~\eqref{ta-variance}.
In particular, the bound Eq.~\eqref{diffusion-bound} is generally not an equality, even if the interactions between the particles are harmonic.
In order to gain some insight on when the bound \eqref{diffusion-bound} is quantitatively useful, we consider the simplest case of two harmonically interacting particles with positions $x$ and $y$,
\begin{subequations}
\begin{align}
\dot{x} &= -\mu \big( k_x x + \kappa (x-y)\big) + \sqrt{2 \mu T} \xi_x \\
\dot{y} &= -\mu \big( k_y y - \kappa (x-y)\big) + \sqrt{2 \mu T} \xi_y ,
\end{align}\label{oscillators}%
\end{subequations}
where $k_x$ and $k_y$ determine the confining force on the individual particles and $\kappa$ the interactions between them.
Note that the variance of the time average of $x$ can always be expressed via the correlation function of the fluctuations of $x$,
\begin{align}
\Av{\Delta \bar{x}^2}_t = \frac{2}{t^2} \int_0^t ds \int_0^s dr \ \Av{\Delta x(s) \Delta x(r)}^\text{eq} .
\end{align}
Since in equilibrium, the correlations only depend on the time difference, the long-time limit of the variance can be further simplified to
\begin{align}
\Av{\Delta \bar{x}^2}_t \simeq \frac{2}{t} \int_0^\infty ds \Av{\Delta x(s) \Delta x(0)}^\text{eq} .
\end{align}
For the above model, this can be evaluated explicitly and yields the bound
\begin{align}
D^* = & \frac{\mu T}{1 + \frac{\kappa^2}{(k_y + \kappa)^2}} .
\end{align}
It is obvious that this is always smaller than the true value of $D = \mu T$ and equal to the latter only if $\kappa = 0$, i.~e.~there is no interaction between the two particles.
However, the bound also becomes accurate for $k_y \gg \kappa$.
This can be understood by considering the additional timescale introduced by coupling the particle $y$ to the particle $x$.
If $k_y$ is small, then the characteristic relaxation time of $y$, $\tau_y = 1/(\mu k_y)$ becomes large; this slow relaxation leads to a likewise slow relaxation of $\av{\Delta \bar{x}^2}$ and thus an underestimation of the diffusion coefficient.
On the other hand, if $k_y$ is large, then the relaxation of $y$ is fast and the relaxation of $\av{\Delta \bar{x}^2}$ is unaffected by the second particle.
Thus, while the model described by Eq.~\eqref{oscillators} itself is not very involved, it yields an important insight on whether the bound \eqref{diffusion-bound} is useful in a given physical situation with interactions.
If the interactions do not have a strong impact on the typical relaxation time of the observed particle, then the estimate $D^*$ of the diffusion coefficient will generally be good.
On the other hand, if the interactions introduce slow modes that strongly influence the relaxation of the observed particle, then the bound \eqref{diffusion-bound} can be much smaller than the actual free-space diffusivity of the particle.

\textit{Derivation of the bound.}
In order to derive the bound \eqref{diffusion-bound} on the diffusion coefficient, we use the fluctuation-response inequality derived in Ref.~\cite{Dec18B}.
There, it was shown that the average response $\delta \av{R}_t$ of a time-integrated stochastic current $R(t)$ to a small perturbation $\delta F$ to the system is bounded from above by the fluctuations $\av{\Delta R^2}_t = \av{R^2}_t - \av{R}_t^2$ of the current times the relative entropy between the path probabilities of the perturbed and unperturbed system,
\begin{align}
\big( \delta \av{R}_t \big)^2 \leq 2 \Av{\Delta R^2}_t \mathbb{S}(\mathbb{P}_{\delta F} \Vert \mathbb{P}) \label{fri} .
\end{align}
In our case, we choose the current to be $R(t) = X_i(t) = \int_0^t ds \ x_i(t)$ and the perturbation to be a small constant force acting on the particle $i$, $\delta F_i$.
The relative entropy can be evaluated explicitly for the dynamics Eq.~\eqref{langevin} and the fluctuation-response inequality reads \cite{Dec18B}
\begin{align}
\bigg(\int_0^t ds \ \delta \av{x_i}_s \bigg)^2 \leq \frac{1}{2} \Av{\Delta X_i^2}_t \frac{\mu \delta F_i^2}{T} t \label{fri-const} .
\end{align}
For long times, the system will settle into a new equilibrium state
\begin{align}
P^\text{eq}_{\delta F}(\bm{x}) = \frac{\exp \big[-\frac{U(\bm{x}) - \delta F_i x_i}{T} \big]}{\int d\bm{x} \ \exp \big[-\frac{U(\bm{x}) - \delta F_i x_i}{T} \big]} ,
\end{align}
with the corresponding shift in the average position
\begin{align}
\delta \av{x_i}^\text{eq} = \frac{\delta F_i}{T} \Av{\Delta x_i^2}^\text{eq} + O(\delta F_i^2) .
\end{align}
Plugging this into Eq.~\eqref{fri-const} and dividing by $t^2$, we obtain
\begin{align}
\frac{1}{t} \Av{\Delta X_i^2}_t \geq 2 \frac{\big(\av{\Delta x_i^2}^\text{eq}\big)^2}{\mu T},
\end{align}
valid for long times.
Noting that $X_i$ is related to the time average by $X_i = t \bar{x}_i$ and identifying the free-space diffusion coefficient $D = \mu T$, this is equivalent to Eq.~\eqref{diffusion-bound}
\begin{align}
D \geq \lim_{t \rightarrow \infty} \frac{2 \big(\av{\Delta x_i^2}^\text{eq}\big)^2}{t \av{\Delta \bar{x}_i^2}_t} .
\end{align}
Note that the measurement of $\bar{x}_i$ is entirely passive; it does not require actually applying the force $\delta F_i$.
Rather, the latter is a virtual probe force that is used only for the derivation of the inequality \eqref{diffusion-bound}.
In Ref.~\cite{Dec18B} it was noted that the form of the fluctuation-response inequality \eqref{fri} is independent of whether the system is over- or underdamped; thus the above discussion and the resulting bound Eq.~\eqref{diffusion-bound} also apply to underdamped systems.
While the bound \eqref{diffusion-bound} holds for the long-time limit, it may be violated if the right-hand side is evaluated at finite times.
However, the fluctuation-response inequality \eqref{fri} also allows us to derive a finite-time version of the bound, by assuming that the system is initially in the modified equilibrium state $P_{\delta F}^\text{eq}$.
Doing so adds another contribution to the relative entropy, which is given by the relative entropy between the two equilibrium distributions $\mathbb{S}(P_{\delta F}^\text{eq} \Vert P^\text{eq})$ \cite{Dec18B}.
For small $\delta F$, this can be evaluated explicitly, $\mathbb{S}(P_{\delta F}^\text{eq} \Vert P^\text{eq}) \simeq
(\delta F_i/T)^2 \av{\Delta x_i^2}^\text{eq}$, and yields the bound 
\begin{align}
D \geq \frac{2 \big(\av{\Delta x_i^2}^\text{eq}\big)^2}{t \av{\Delta \bar{x}_i^2}_t} - 2 \frac{\av{\Delta x_i^2}^\text{eq}}{t} \label{diffusion-bound-finite} .
\end{align}
This is converges to the bound \eqref{diffusion-bound} in the long-time limit.
While Eq.~\eqref{diffusion-bound-finite} obviously does not yield a useful bound at short times, it shows that corrections to the bound Eq.~\eqref{diffusion-bound} are of order $1/t$ at finite times.

\textit{Explicit demonstration.}
The bound \eqref{diffusion-bound} always holds irrespective of the trapping potential, interactions and whether the description of the system is over- or underdamped. 
For a realistic physical situation consisting of many interacting particles in a generally anharmonic trapping potential an obvious question is: How useful is the estimate $D^*$ for determining the true free-space diffusivity $D$?
To shed some light on this issue, we perform a numerical simulation of trapped, interacting particles and determine $D^*$ from the measurement of the trajectories in order to compare it to the known value of $D$.
Specifically, we consider $M = 20$ identical spherical particles of radius $\rho$ and mass $m$.
The particles are trapped inside a three-dimensional trap with size $L$ and depth $U_\text{t}$
\begin{align}
U^\text{trap}(\bm{x}_i) = U_\text{T} \bigg( \frac{|\bm{x}_i|^2}{2 L^2} - \frac{|\bm{x}_i|^4}{8 L^4} + \frac{|\bm{x}_i|^6}{48 L^6} \bigg) \label{trapping-pot} .
\end{align}
This trapping potential can be considered as the expansion of a Gaussian trap,
%\begin{align}
%U^\text{trap}(\bm{x}_i) = -U_\text{T} \bigg( e^{-\frac{|\bm{x}_i|^2}{2 L^2}} - 1 \bigg) \label{trap-gaussian},
%\end{align}
which is frequently encountered in experimental situations, e.~g.~for optical trapping.
The form Eq.~\eqref{trapping-pot} takes into account anharmonic corrections while avoiding the problem that a Gaussian trap is not a confining potential.
The particles further interact via a short-range repulsive pair potential with $r_{i j} = |\bm{x}_i - \bm{x}_j|$
\begin{align}
U^\text{int}(r_{i j}) = \left\lbrace \begin{array}{ll}
U_\text{I} \Big( \frac{1}{2} \big(\frac{2\rho}{r_{ij}}\big)^4 - \big(\frac{2\rho}{r_{ij}}\big)^2 + \frac{1}{2} \Big), &r_{ij} < 2\rho \\[2ex]
0, &r_{ij} \geq 2\rho .
\end{array} \right.
\end{align}
This potential is similar to a truncated Lennard-Jones potential, however, we choose to use a combination of $r^{-4}$ and $r^{-2}$ terms, instead of $r^{-12}$ and $r^{-6}$, as it allows using a larger timestep in the numerical simulations.
The interaction between the particles and the surrounding medium is incorporated via Stokes friction and thermal noise, resulting in the set of underdamped Langevin equations
\begin{align}
m \dot{\bm{v}}_i = -\frac{\bm{v}_i}{\mu}  - \bm{\nabla}_i \Big( U^\text{trap}(\bm{x}_i) + \sum_{j \neq i} U^\text{int}(r_{i j}) \Big) + \sqrt{\frac{2 T}{\mu}} \bm{\xi}_i \label{langevin-int}
\end{align}
where $\bm{r}_i = (x,y,z)_i$ and $\bm{v}_i=(v_{x},v_{y},v_{z})_i$ are the position and velocity of the $i$-th particle and $\bm{\nabla}_i$ denotes the gradient with respect to $\bm{x}_i$.
As before $\mu$ is the particle mobility and $T$ the temperature.
This example features three effects that may potentially spoil the accuracy of the estimate $D^*$: anharmonic trapping, interactions and underdamped motion.
We perform Langevin simulations of Eq.~\eqref{langevin-int}, recording for each particle the position and time-averaged position.
Then, averaging over the particles, we compute the respective variances and the bound \eqref{diffusion-bound}.
To limit the effect of statistical fluctuations, we repeat this procedure a total of $600$ times and also average over these realizations.
Note that the averaging over different realizations is only done for numerical convenience, as it allows easy usage of parallel computation.
Since the measurement is performed entirely in equilibrium, one may equally well observe a single realization for a longer time, and divide this long observation into independent segments over which averaging can be performed.
To start, we take non-interacting particles in a purely harmonic confinement with $U_\text{T} = 5$ and $L = 5$.
For $T = 1$, $\mu = 1$ and $m = 1$, we obtain $D^*/D = 1.00 \pm 0.02$, where the uncertainty is the temporal variance of the right-hand side of Eq.~\eqref{diffusion-bound} around its long-time-averaged value.
Thus, for non-interacting harmonically trapped particles under moderate damping, the estimate of the diffusivity is excellent.
Note that this remains valid in the heavily underdamped regime $\mu = 10$, where we obtain $D^*/D = 0.99 \pm 0.02$.
For a harmonic confinement, the position and velocity degrees of freedom decouple in the long-time limit independent of the value of $\mu$ and thus we have $D^* = D$ even at low damping.
Taking into account the anharmonic terms in the potential, we arrive at $D^*/D = 0.98 \pm 0.02$ at $\mu = 1$ and $D^*/D = 0.93 \pm 0.01$ at $\mu = 10$.
While for moderately strong damping, we again obtain an excellent estimate of $D$, the anharmonic terms induce a coupling between the statistics of position and velocity and the slow relaxation of the velocity for low damping reduces the accuracy of the estimate.
Finally, we switch on the interactions with $U_\text{I} = 2.5$ and $\rho = 0.8$, corresponding to an occupied volume fraction of roughly $(U_\text{t}/T)^{3/2} (\rho/L)^3 = 0.05$.
For moderate damping $\mu = 1$, we still find a useful estimate of the diffusion coefficient $D^*/D = 0.90 \pm 0.01$.
Even though we have an anharmonic confinement and moderately strong interactions between the particles, the fact that the thermal relaxation time $\tau_\text{rel} = m \mu = 1$ is about one order of magnitude smaller than the average time between between collisions $\tau_\text{col} \approx 13$ means that the relaxation of the time-average $\bar{x}$ is still mostly governed by the diffusive dynamics.
By contrast, for low damping $\mu = 10$, the relaxation and collision time are of the same order and we find $D^*/D = 0.50 \pm 0.01$, which has still the right order of magnitude but is no longer quantitatively correct.
Reducing the particle radius to $\rho = 0.4$ and thus the rate at which interactions occur to $\tau_\text{col} \approx 47$, the estimate accordingly improves to $D^*/D = 0.97 \pm 0.02$ for $\mu = 1$ and $D^*/D = 0.85 \pm 0.02$ for $\mu = 10$.
From the results of our simulations, which are summarized in Tab.~\ref{tab1}, we conclude that even for interacting particles in an anharmonic trapping potential, the quantity $D^*$, Eq.~\eqref{diffusion-bound}, can provide a good approximation of the free-space diffusion coefficient, as long as the interaction with the heat bath dominates relaxation processes in the system.

\begin{table}
\begin{tabular}{|c|c|c|}
\hline 
setting & moderate damping & low damping \\ 
\hline 
$\begin{array}{cc}
\text{harmonic} \\[-1ex] \text{non-interacting}
\end{array}$ & 1.00 & 0.99 \\ 
\hline 
$\begin{array}{cc}
\text{anharmonic} \\[-1ex] \text{non-interacting}
\end{array}$ & 0.98 & 0.93 \\ 
\hline 
$\begin{array}{cc}
\text{anharmonic} \\[-1ex] \text{low density}
\end{array}$ & 0.97 & 0.85 \\ 
\hline 
$\begin{array}{cc}
\text{anharmonic} \\[-1ex] \text{high density}
\end{array}$ & 0.90 & 0.50 \\ 
\hline 
\end{tabular} 
\caption{Ratio of the estimate $D^*$, Eq.~\eqref{diffusion-bound}, and the true value of the diffusivity $D$, obtained for different physical situations from numerical simulations of Eq.~\eqref{langevin-int}. \label{tab1} }
\end{table}

\textit{Discussion.}
We have found a possibility to estimate the free-space diffusion coefficient by observing the trajectories of confined particles in an equilibrium state.
The estimate is exact for non-interacting particles in a harmonic confinement, but remains accurate in the presence weak anharmonic effects and interactions between the particles.
Even when the validity of these assumptions for a given experimental situation cannot easily be assessed, the obtained value of $D^*$ is guaranteed to provide a lower bound on the true diffusion coefficient $D$.

We anticipate our estimate to be useful in physical situations where a direct measurement of the diffusion coefficient is not possible, in particular for small systems, where diffusive behavior is only observable on short timescales.
However, even if a direct observation of diffusion is possible, observing the motion of trapped particles may be more convenient from an experimental point of view, since small particles may be difficult to track over long distances.
Trapping remedies this issue and allows for arbitrarily long observation times.
Moreover, the trapping potential can be adjusted to exclude harmonic effects and interactions, allowing to estimate the diffusion coefficient with good accuracy.

\begin{acknowledgments}
\textbf{Acknowledgments.} This work was supported by the World Premier International Research Center Initiative (WPI), MEXT, Japan.
\end{acknowledgments}

\bibliography{bib}

\end{document}